\documentclass[amssymb,prb,twocolumn,showpacs]{revtex4}
\usepackage{epsfig}
\usepackage{dcolumn}
\usepackage{amsmath}
\begin{document}
\title{\bf Effect of frequency and temperature on  microwave-induced
magnetoresistance oscillations in two-dimensional electron systems.}
\author{Jes\'us I\~narrea$^{1,2}$}
 \affiliation {$^1$Escuela Polit\'ecnica
Superior,Universidad Carlos III,Leganes,Madrid,Spain and  \\
$^2$Unidad Asociada al Instituto de Ciencia de Materiales, CSIC,
Cantoblanco,Madrid,28049,Spain.}
\date{\today}
%%%%%%%%%%%%%%%%%%%%%%%%%%%%%%%%%%%%%%%%%%%%%%%%%%%%%%%%%%%%%%%%%%
%%%%%%%%%%%%
\begin{abstract}
Experimental results on microwave-induced magnetoresistance
oscillation in two-dimensional electron systems show a similar
behavior of these systems regarding temperature and microwave
frequency. It is found that these oscillations tend to quench when
frequency or temperature increase, approaching magnetoresistance to
the response of the dark system. In this work we show that this
experimental behavior can be addressed on the same theoretical
basis. Microwave radiation forces the electron orbits to move back
and forth being damped by interaction with the lattice. We show that
this damping depends dramatically on microwave frequency and also on
temperature. An increase in frequency or temperature gives rise to
an increase in the lattice damping producing eventually a quenching
effect in the magnetoresistance oscillations.

\end{abstract}
%%%%%%%%%%%%%%%%%%%%%%%%%%%%%%%%%%%%%%%%%%%%%%%%%%%%%%%%%%%%%%%%%%
%%%%%%%%%%%%
\maketitle The study of the effects that radiation can produce on
nano-devices, such as two-dimensional electron systems (2DES), is
attracting considerable attention since the last two decades both
from theoretical and experimental sides\cite{ina}. Important and
unusual properties have been discovered when these systems are
subjected to different external potentials. In particular
microwave-induced magentoresistance oscillations (MIRO) and Zero
Resistance States (ZRS), are being a subject of intense
investigation since its recent
discovery\cite{mani,zudov,studenikin}. Since then, experimental
results are being produced in a continuous basis which challenge the
available theoretical
\begin{figure} \centering\epsfxsize=3.5in
\epsfysize=4.0in \epsffile{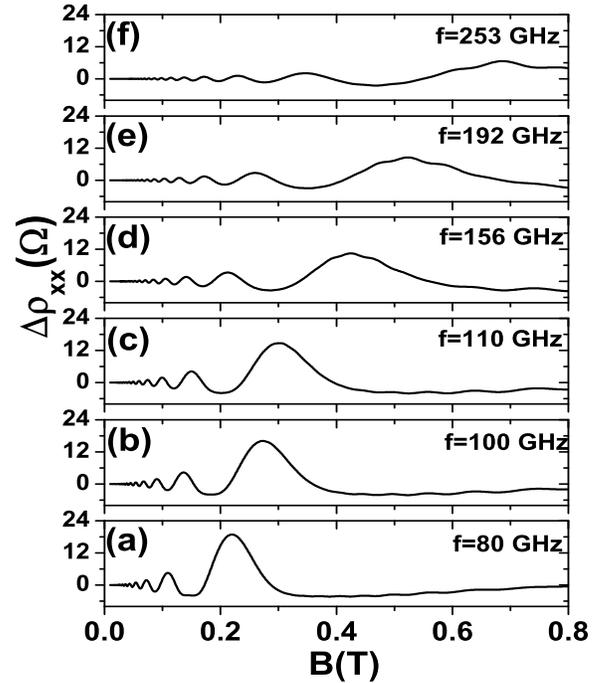} \caption{ Calculated
magnetoresistance difference between radiation and dark
situations:$\Delta \rho_{xx}=\rho_{xx}^{MW}-\rho_{xx}^{Dark}$ as a
function of magnetic field $B$ for different MW frequencies from
$80$ to $253$ GHZ. $T_{L}=1 K$. }
\end{figure}
models\cite{ina2,girvin,dietel,lei,ryzhii,rivera,shi,andreev,mirlin}.
Among all of those experimental outcomes we can cite for instance,
temperature dependence of MIRO\cite{mani,zudov,ina2}, absolute
negative conductivity \cite{willett,ina3,bykov}, bichromatic
microwave (MW) exictation\cite{zudov2,ina4}, polarization immunity
of magnetoresistance ($\rho_{xx}$)\cite{smet,ina5}, effect of
in-plane magnetic fields\cite{yang}. More recently another striking
experimental result by Studenikin et al.,\cite{studenikin2} has
joined the group. They demonstrate that at high MW frequencies
(above $120$ GHz) MIRO start to quench and for further increasing
frequencies,  MIRO practically  disappear. They also show
experimental results on temperature dependence of MIRO amplitude
indicating that at increasing lattice temperature the oscillations
amplitude dramatically decrease. This behavior was already obtained
by the seminal experiments on MIRO and
ZRS\cite{mani,zudov,studenikin}. For both cases, MW frequency and
temperature, the system tends to recover the dark $\rho_{xx}$
response.

In this letter we present a common theoretical approach to explain
the quanching effect of both, MW frequency and temperature, on MIRO.
The starting point is the effect of MW radiation on 2DES, which is
treated with the MW driving Larmor orbits model\cite{ina2,ina3}.
According to it,  MW radiation forces electronic orbits to move
periodically back and forth at MW frequency. As a result, the whole
two-dimensional electron gas (2DEG) moves harmonically in the
current direction. In this periodic movement electrons in their
orbits (quantum oscillators) interact with the lattice being damped
and emiting acoustic phonons. If we increase the MW frequency
leaving fixed the lattice temperature ($T_{L}$), the average number
of interactions between electrons and lattice ions will be larger.
Thus, we obtain and increasing damping of the MW-driven oscillating
motion. This is finally reflected in the decreasing amplitude of
MIRO. On the other hand if we increase the $T_{L}$ leaving fixed the
MW frequency $w$, the average electrons-lattice interactions will be
larger producing more damping and smaller $\rho_{xx}$ oscillations.
Thus, the quenching of MIRO by high $w$ or $T_{L}$ would be based on
the same physical effect.

\begin{figure}
\centering\epsfxsize=3.0in \epsfysize=3.0in \epsffile{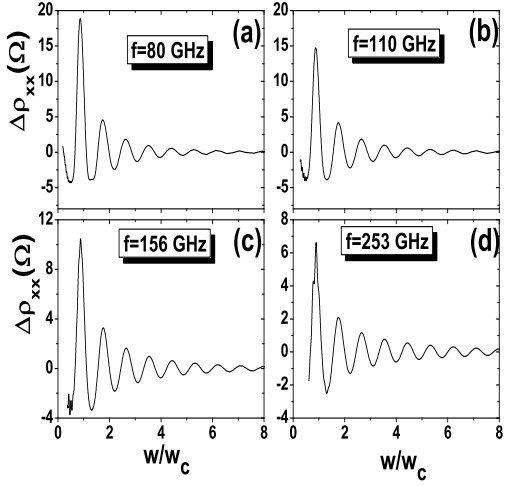}
\caption{(Calculated $\rho_{xx}$ difference for different
frequencies plotted as a functions of $w/w_{c}$. $T_{L}=1K$.}
\end{figure}

We have recently\cite{ina3} proposed a model to explain the
$\rho_{xx}$ behavior of a 2DEG at low $B$ and under MW radiation.
According to it, due to the MW radiation, center position of
electronic orbits are not fixed, but they oscillate back and forth
harmonically with $w$. The amplitude $A$ for these harmonic
oscillations is given by:
\begin{equation}
A=\frac{e E_{o}}{m^{*}\sqrt{(w_{c}^{2}-w^{2})^{2}+\gamma^{4}}}
\end{equation}
where $e$ is the electron charge, $w_{c}$ the cyclotron frequency,
$E_{0}$ the intensity for the MW field and $\gamma$ is a damping
parameter due to interaction with the lattice. Now we introduce the
scattering suffered by the electrons due to charged impurities
randomly distributed in the sample\cite{ridley,ina2,ina3}. Firstly
we calculate the electron-charged impurity scattering rate $1/\tau$,
and secondly we find the average effective distance advanced by the
electron in every scattering jump: $\Delta X^{MW}=\Delta X^{0}+
A\cos w\tau$, where $\Delta X^{0}$ is the effective distance
advanced when there is no MW field present. Finally the longitudinal
conductivity $\sigma_{xx}$ can be calculated: $\sigma_{xx}\propto
\int dE \frac{\Delta X^{MW}}{\tau}(f_{i}-f_{f})$,  being $f_{i}$ and
$f_{f}$ the corresponding distribution functions for the initial and
final Landau states respectively and $E$ energy. To obtain
$\rho_{xx}$ we use the relation
$\rho_{xx}=\frac{\sigma_{xx}}{\sigma_{xx}^{2}+\sigma_{xy}^{2}}
\simeq\frac{\sigma_{xx}}{\sigma_{xy}^{2}}$, where
$\sigma_{xy}\simeq\frac{n_{i}e}{B}$ and $\sigma_{xx}\ll\sigma_{xy}$.
Therefor, $\rho_{xx}$ is proportional to the MW-induced oscillation
amplitude of the electronic orbits center: $\rho_{xx}\propto A \cos
w\tau$.

At this point, we introduce a microscopic model which allows us to
obtain $\gamma$ and its dependence on $w$ and $T_{L}$ as follows.
Following Ando and other authors\cite{ando}, we propose the next
expression for the electron-acoustic phonons scattering rate valid
at low $T_{L}$:
\begin{equation}
\frac{1}{\tau_{ac}}=\frac{m^{*}\Xi_{ac}^{2}k_{B}T_{L}}{\hbar^{3}\rho
u_{l}^{2}<z>}
\end{equation}
where $\Xi_{ac}$ is the acoustic deformation potential, $\rho$ the
mass density, $u_{l}$ the sound velocity and $<z>$ is the effective
layer thickness. However $\frac{1}{\tau_{ac}}$ is not yet the final
expression for $\gamma$. First, we have to multiply
$\frac{1}{\tau_{ac}}$ by the average number of times that an
electron in its orbit can interact effectively with the lattice
ions.  We can approximate this number  dividing the length an
electron runs in its Landau state by the lattice parameter of GaAs,
($a_{GaAs}$): $2\pi R_{c}/a_{GaAs}$, $R_{c}$ being the cyclotron
radius for the corresponding orbit. Secondly, being $\gamma$ a
dissipative term, it is directly proportional to the MW energy and
inversely proportional to the emitted acoustic phonon energy. Thus,
we have finally to multiply by the ratio: $\frac{w}{w_{ac}}$,
$w_{ac}$ being the average frequency of the emitted acoustic
phonons\cite{kent}. Finally we obtain an expression for $\gamma$:
\begin{equation}
\gamma=\frac{1}{\tau_{ac}} \times \frac{2\pi R_{c}}{a_{GaAs}} \times
\frac{w}{w_{ac}}
\end{equation}
 With the
experimental parameters we have at
hand\cite{mani,zudov,studenikin,kent} and for an average magnetic
field it is straightforward to obtain a direct relation between
$\gamma$ and $T_{L}$ and $w$, resulting in a linear dependence:
\begin{equation}
\gamma \propto T_{L} \times w
\end{equation}
 Now is possible to go further
and calculate the variation of $\rho_{xx}$ with $T_{L}$ and $w$.
\begin{figure}
\centering\epsfxsize=3.0in \epsfysize=4.0in \epsffile{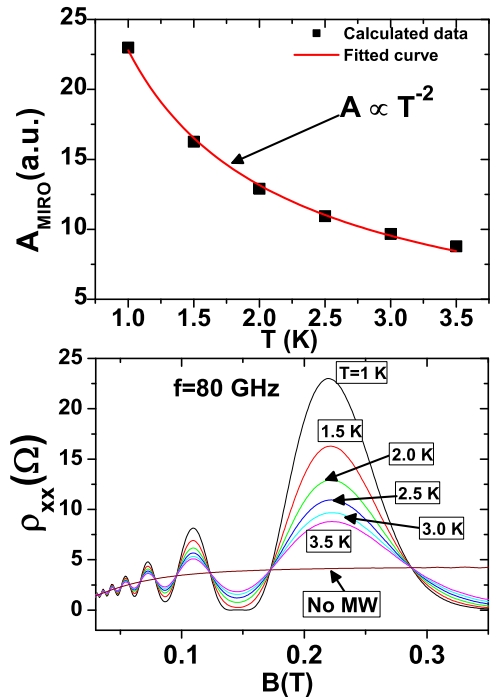}
\caption{(Color on line). Bottom panel: calculated $\rho_{xx}$ as a
function of $B$ for different $T_{L}$ from $1$K to $3.5$K.
$\rho_{xx}$ presents decreasing oscillations for increasing $T_{L}$
at a fixed $w=80$ GHz. $T_{L}=1$ K. Bottom panel: $T_{L}$ dependence
of MIRO amplitude in arbitrary units determined by fitting
calculated $\rho_{xx}$ results of the bottom panel. Calculated data
have obtained from the largest $\rho_{xx}$ peak (around $B=0.2$ T}
\end{figure}

In Fig. 1, we present calculated  $\rho_{xx}$ difference (between
radiation and dark situations:$\Delta
\rho_{xx}=\rho_{xx}^{MW}-\rho_{xx}^{Dark}$)  as a function $B$ for
different MW frequencies from $80$ to $253$ GHZ. From Fig.1a to 1f
the $\rho_{xx}$ evolution shows the quenching of MIRO being clearly
visible from $w=110$ GHz. At $w=253$ GHz are hardly visible. The
explanation can be readily obtained according to our theory. As we
said above, $\rho_{xx}$ is proportional to $A$:
\begin{equation}
\rho_{xx} \propto\frac{e
E_{o}}{m^{*}\sqrt{(w_{c}^{2}-w^{2})^{2}+\gamma^{4}}} \cos w\tau
\end{equation}
From equations (4) and (5), we can deduce that higher values for $w$
mean larger $\gamma$,  finally reflected in smaller $\rho_{xx}$.
Eventually for large enough $w$, $\rho_{xx}$ will be practically
wiped out. In Fig. 2, we present calculated $\Delta \rho_{xx}$ for
different frequencies plotted as a function of $w/w_{c}$. Calculated
results of Figs. 1 and 2 are in reasonable agreement with
experiments\cite{studenikin2}.

In Fig. 3  bottom panel, we show calculated $\rho_{xx}$ as a
function of $B$ for different $T_{L}$ from $1$K to $3.5$K.
$\rho_{xx}$ presents decreasing oscillations for increasing $T_{L}$
at a fixed $w=80$ GHz. In the top panel we represent $T_{L}$
dependence of MIRO amplitude in arbitrary units determined by
fitting calculated $\rho_{xx}$ results of the bottom panel.
Calculated data have obtained from the largest $\rho_{xx}$ peak
(around $B=0.2$ T). According to this fitting MIRO amplitude depends
roughly on $T_{L}$ as $A_{MIRO}\propto T_{L}^{-2}$. This behavior is
explained similarly as in the $w$ case following equations (4) and
(5) including the $A_{MIRO}$ dependence on $T_{L}$.

This work has been supported by the MCYT (Spain) under grant
MAT2005-0644, by the Ram\'on y Cajal program and  by the EU Human
Potential Programme: HPRN-CT-2000-00144.

\end{document}